\documentclass[aps, pre, twocolumn]{revtex4}

\usepackage{graphicx}
\usepackage{amsmath}
\usepackage{amssymb}
\usepackage{wasysym}

\newcommand{\mPIV}{$\mu$-PIV~}
\newcommand{\unit}[1]{\ensuremath{\, \mathrm{#1}}}

\begin{document}
\title{Quantifying effective slip length over micropatterned hydrophobic surfaces}

\author{Peichun Tsai$^{1}$, Alisia M. Peters$^{2}$, Christophe Pirat$^{1, 3}$, Matthias Wessling$^{2}$, Rob G. H. Lammertink$^{2}$, and Detlef Lohse$^{1}$}

\address{$^{1}$ Physics of Fluid Group, Department of Applied Physics, $^{2}$ Membrane Science and Technology Group, Department of Chemical Engineering, University of Twente, P.O. Box 217, 7500 AE, Enschede, The Netherlands, $^{3}$Present address: Laboratoire de Physique de la Mati\`ere Condens\'ee et Nanostructures, Universit\'e de Lyon; Univ. Lyon I, CNRS, UMR 5586, 69622 Villeurbanne, France}

\date{\today}

\begin{abstract}
We employ micro-particle image velocimetry ($\mu$-PIV) to investigate laminar micro-flows in hydrophobic microstructured channels, in particular the slip length.  These microchannels consist of longitudinal micro-grooves, which can trap air and prompt a shear-free boundary condition and thus slippage enhancement.   
Our measurements reveal an increase of the slip length when the width of the micro-grooves is enlarged.  The result of the slip length is smaller than the analytical prediction by Philip et al.~\cite{JPhilip} for an infinitely large and textured channel comprised of alternating shear-free and no-slip boundary conditions.  The smaller slip length (as compared to the prediction) can be attributed to the confinement of the microchannel and the bending of the meniscus (liquid-gas interface).  Our experimental studies suggest that the curvature of the meniscus plays an important role in microflows over hydrophobic micro-ridges.   
\end{abstract}

\maketitle
\section{Introduction}

Accurately modeling hydrodynamic boundary condition is essential in fluid dynamics. 
The notion of the no-slip boundary condition at a solid-liquid interface has been successfully employed for macroscopic flows for more than a century; while its validation for micro- and nano- flows has been under an intensive debate over the last two decades~\cite{Vinogradova:1999p486, ELauga_noslip_BC_2007, Bocquet_2007}. 
Recent measurements of slip length for flat hydrophobic surfaces with different experimental techniques are discrepant, presenting a wide range of slip length scales from $nm$ to a few $\mu m$~\cite{Meinhart_PhysFluids_2002,PTabeling_PRE_2005,Craig_review_Rep_Prog_Phys05,SFA_cottin-bizonne_PRL_2005}, though recently it became clearer and clearer that on a flat surface a slip length in the range of a few nanometers is more realistic~\cite{Bocquet_PRL2006,SFA_cottin-bizonne_PRL_2005, Bocquet_PRL2006_2}.  In any case, careful and systematic experimental means are necessary to clarify the correct boundary condition at micro- and nano- scale to validate the existing theoretical models and numerical simulations.

The combination of chemical hydrophobicity and physical roughness can allow for air-liquid interfaces upon solid surfaces~\cite{deGennes_book, SGranick_PRL06}.  When this idea is applied to microfluidic devices, slippage can be promoted due to the shear-free boundary condition at the air-liquid menisci, and hence drag reduction can be achieved, as investigated with theory~\cite{JPhilip,Lauga_JFM2003,Mauro}, simulation~\cite{LBocquet_EuroPhysJE04,mauro_JFM2006}, and experiment~\cite{JRothstein_PhysFluids_2004,JRothstein_PhysFluids_2005,Byun_Phys_Fluids_2008}. 
Both the surface wettability and roughness can affect the hydrodynamic boundary condition~\cite{ECharlaix_2007}.     
Moreover, the shapes of air-liquid menisci have recently been demonstrated to change the slippage condition of hydrophobic walls~\cite{SFA_cottin-bizonne_PRL_2005, ECharlaix_2007}.

In this paper, hydrophobic polymer channels were fabricated and patterned with longitudinal grooves of height $h$ and width $w$ at micron scale. Air can be trapped between the micro-grooves when water steadily flows through the channel, providing that the local liquid pressure is not large.  We conducted a series of experiments of varying geometric parameters to investigate the geometric effect of the air-filled micro-ridges on the boundary slip.  
Two different polymeric surfaces--one smooth and dense and the other rough and porous-- were  used to study the influence of the wall structure on the slippage.  Dense and porous materials are fundamentally different in structure which profoundly affects the mass transport of fluids. Particularly, porous materials are of interest in membrane science and technology due to their potential applications as catalyst supports. 
In addition, we explored the impact of the shape of meniscus and the finite size effect of the confined micro-channels on the slip length.

\begin{figure}[b]
\begin{center}
\includegraphics[width=3.6in]{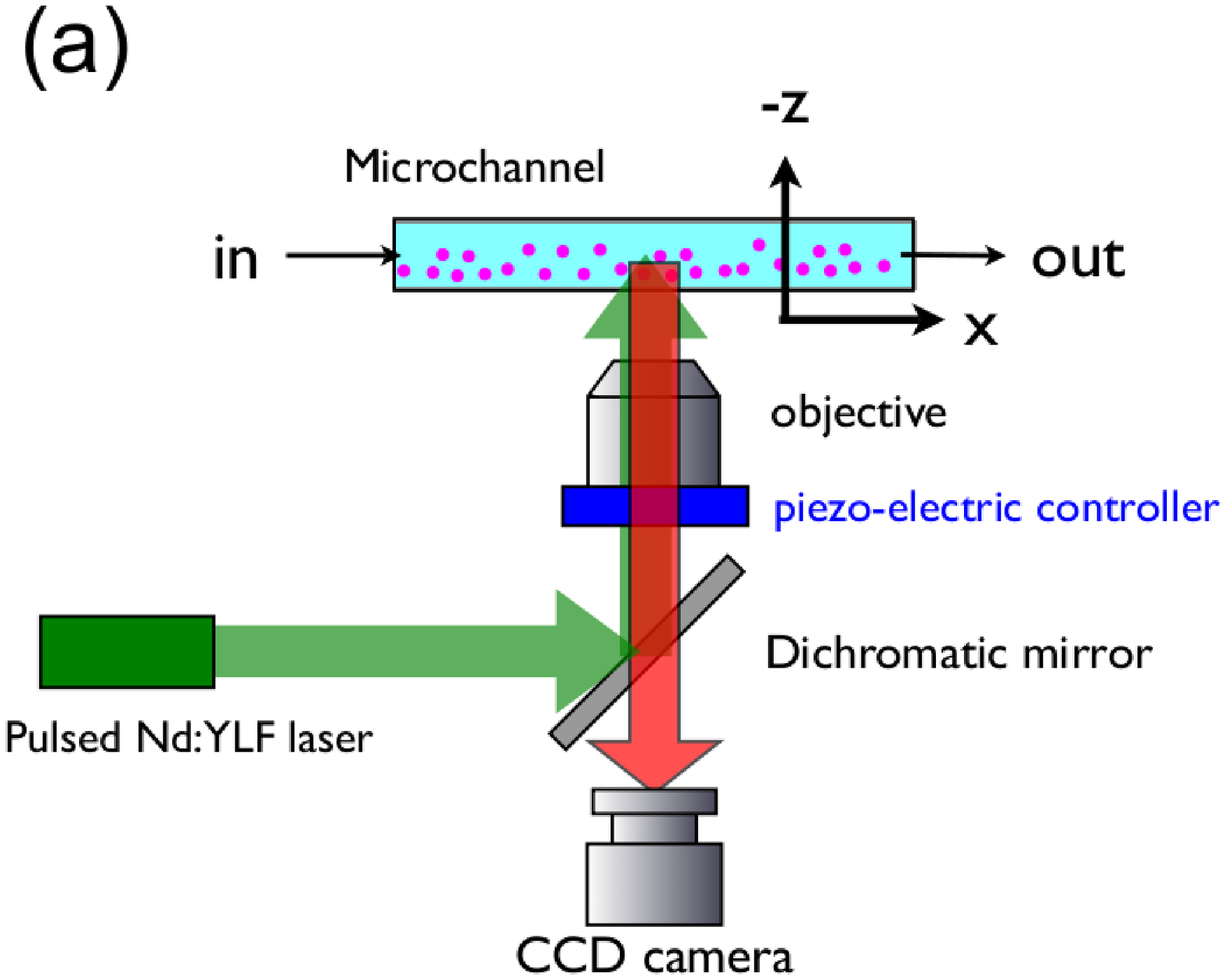}
\includegraphics[width=3.0in]{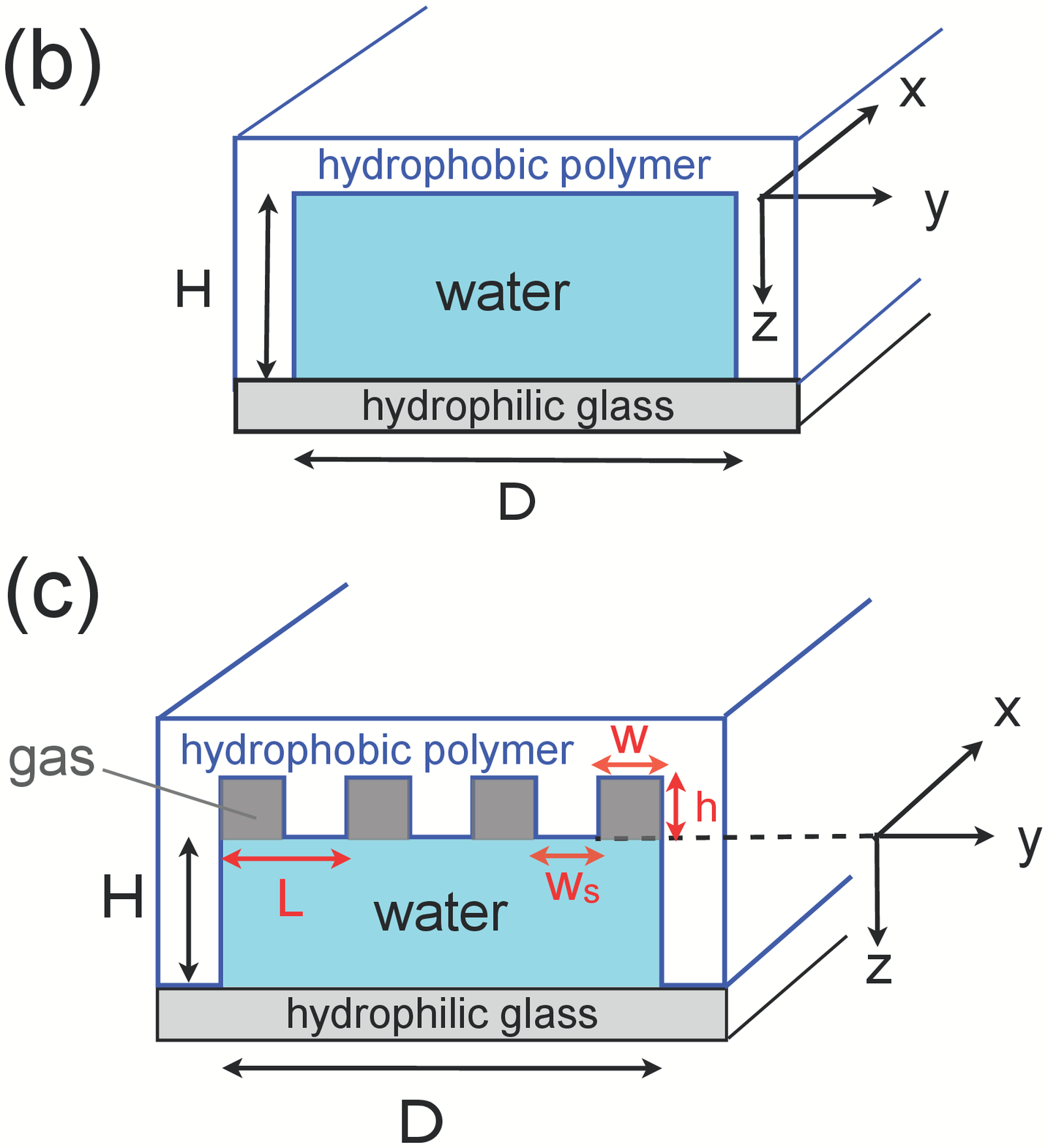}
\caption{\label{setup}Schematic diagrams of (a) the micro-PIV setup, (b) the smooth rectangular microchannel, and (c) the microstructured channel, showing the trapped air inside the grooves when water flows through the channel.
}
\end{center}
\end{figure}

Fig.~\ref{setup}a schematically shows our $\mu-$PIV setup and its working principle. Image pairs of the seeding particles in a microchannel under a constant water flow rate are observed with properly delated double-pulsed laser illumination via an objective and recorded by a CCD camera. The scanning plane of the laser illumination is controlled by a piezo-electric controller to investigate the particle motions in the $x-y$ plane at the different channel heights $z$. The cross-correlation of these captured image pairs allows us to quantify the flow field.   Fig.~\ref{setup} b and c illustrate the structures of our employed rectangular and patterned microchannels, which are comprised of a hydrophobic polymer and a thin enclosure of transparent (hydrophilic) glass.

\section{Experimental Section}

The general experimental procedure includes applying a steady water flow rate which ranges from $0.1~\unit{\mu L/min}$ to $0.5~\unit{\mu L/min}$ (PHD 2000 Infusion syringe pump, Harvard Apparatus) through a microchannel.  Thus, a steady laminar flow develops and the velocity profile is obtained with the \mPIV technique.

\subsection{Preparation of hydrophobic microchannels}

\begin{figure}
\begin{center}
\includegraphics[width=1.68in]{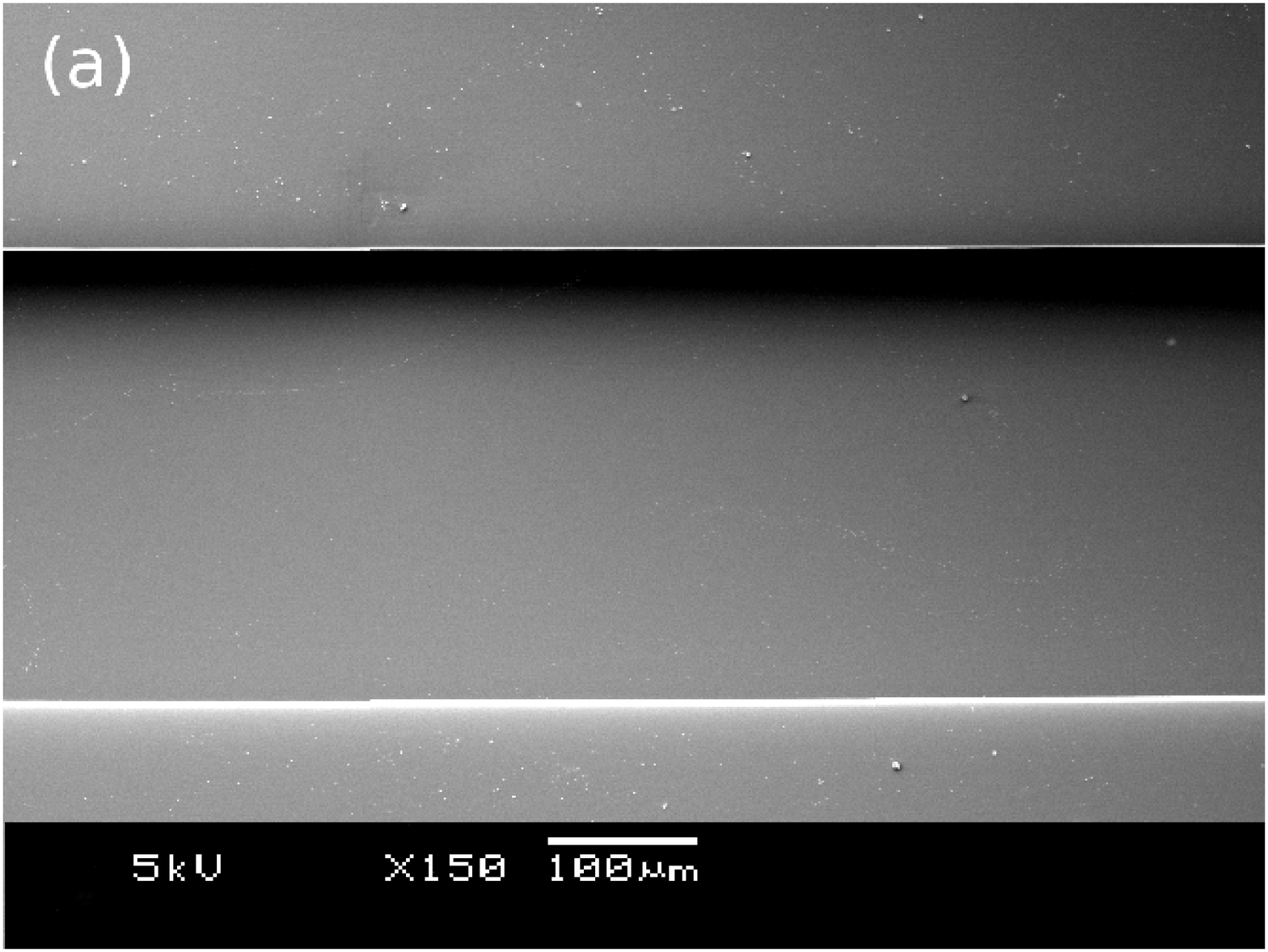}
\includegraphics[width=1.68in]{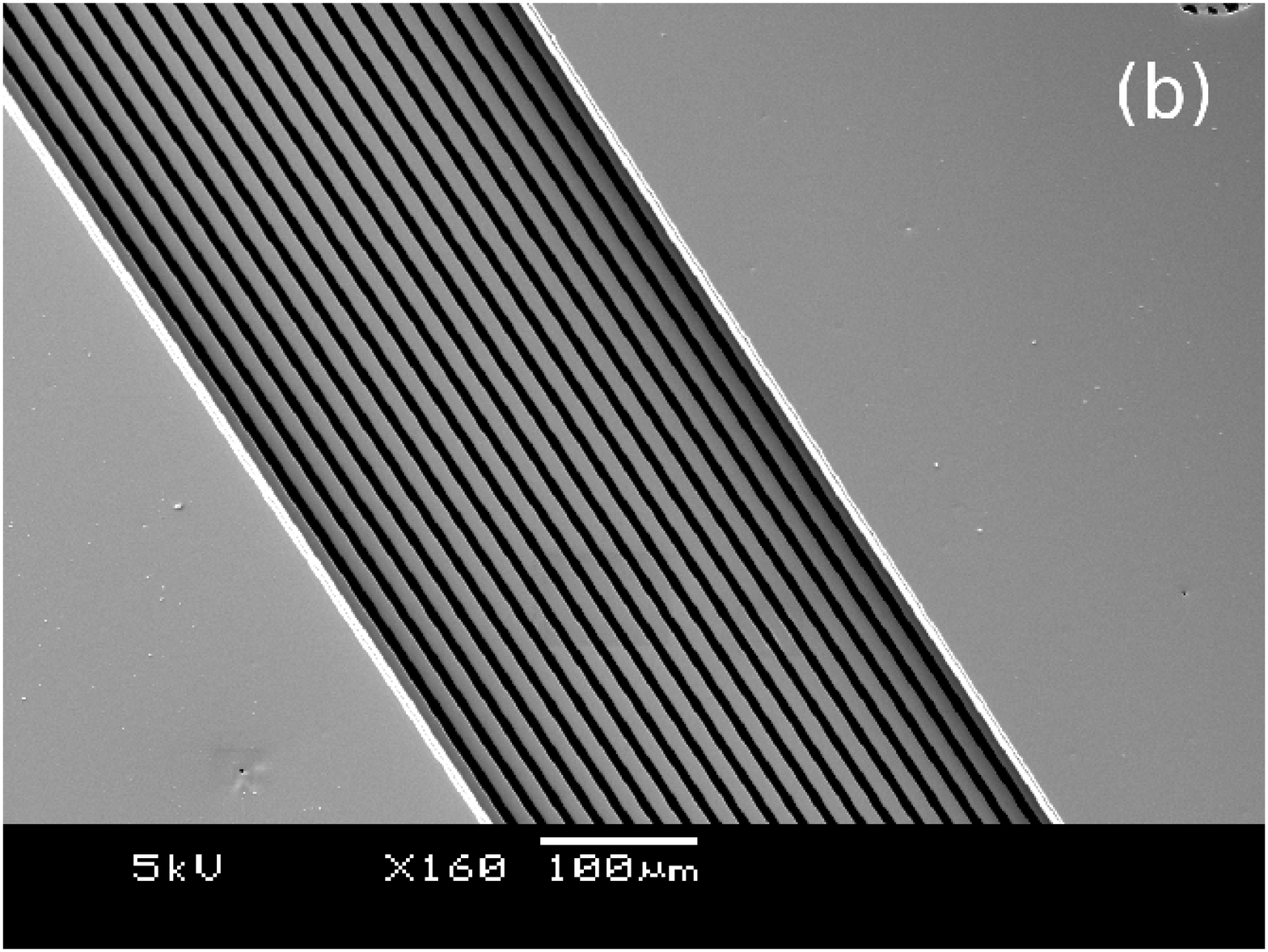}
\includegraphics[width=1.68in]{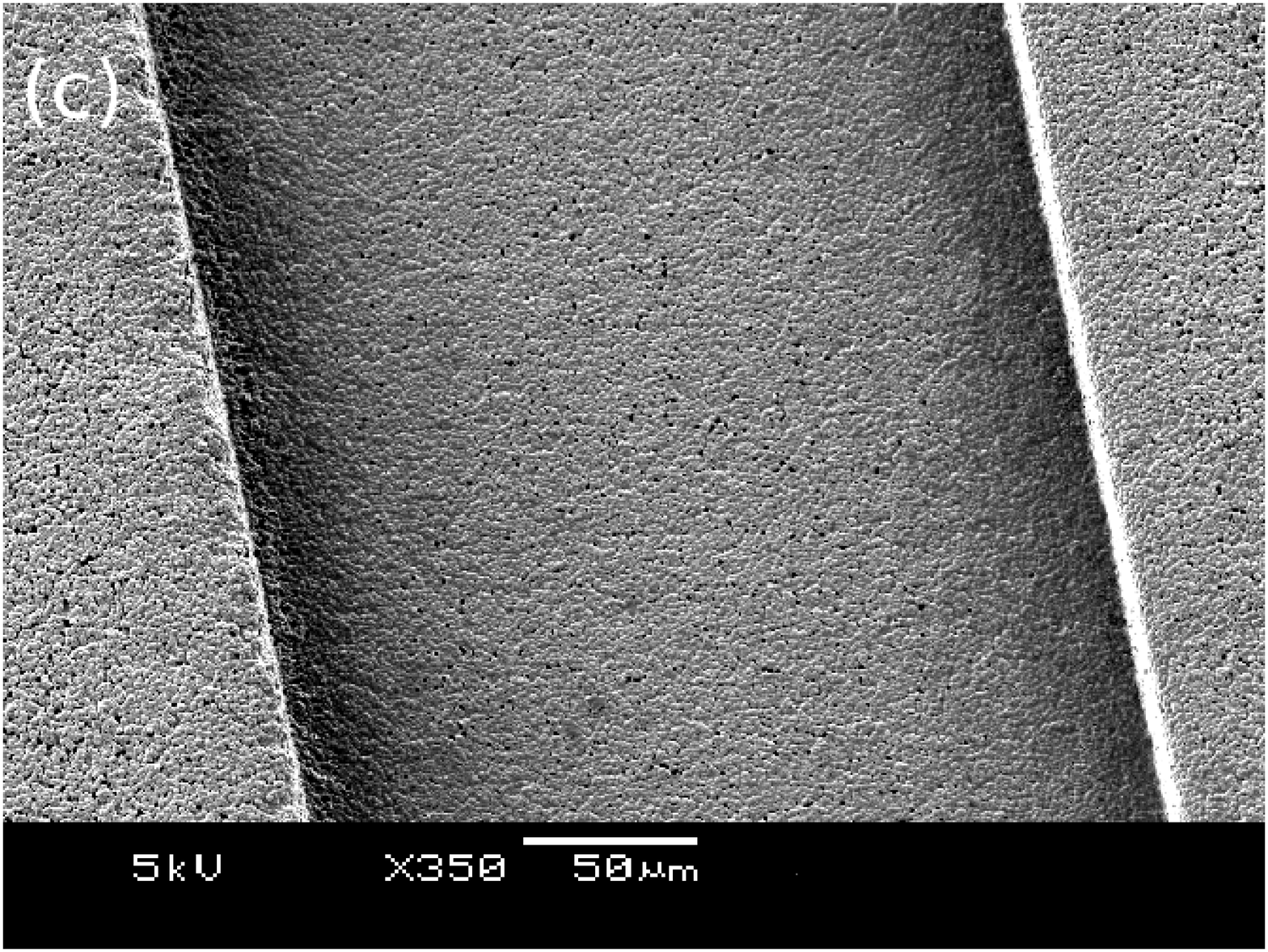}
\includegraphics[width=1.68in]{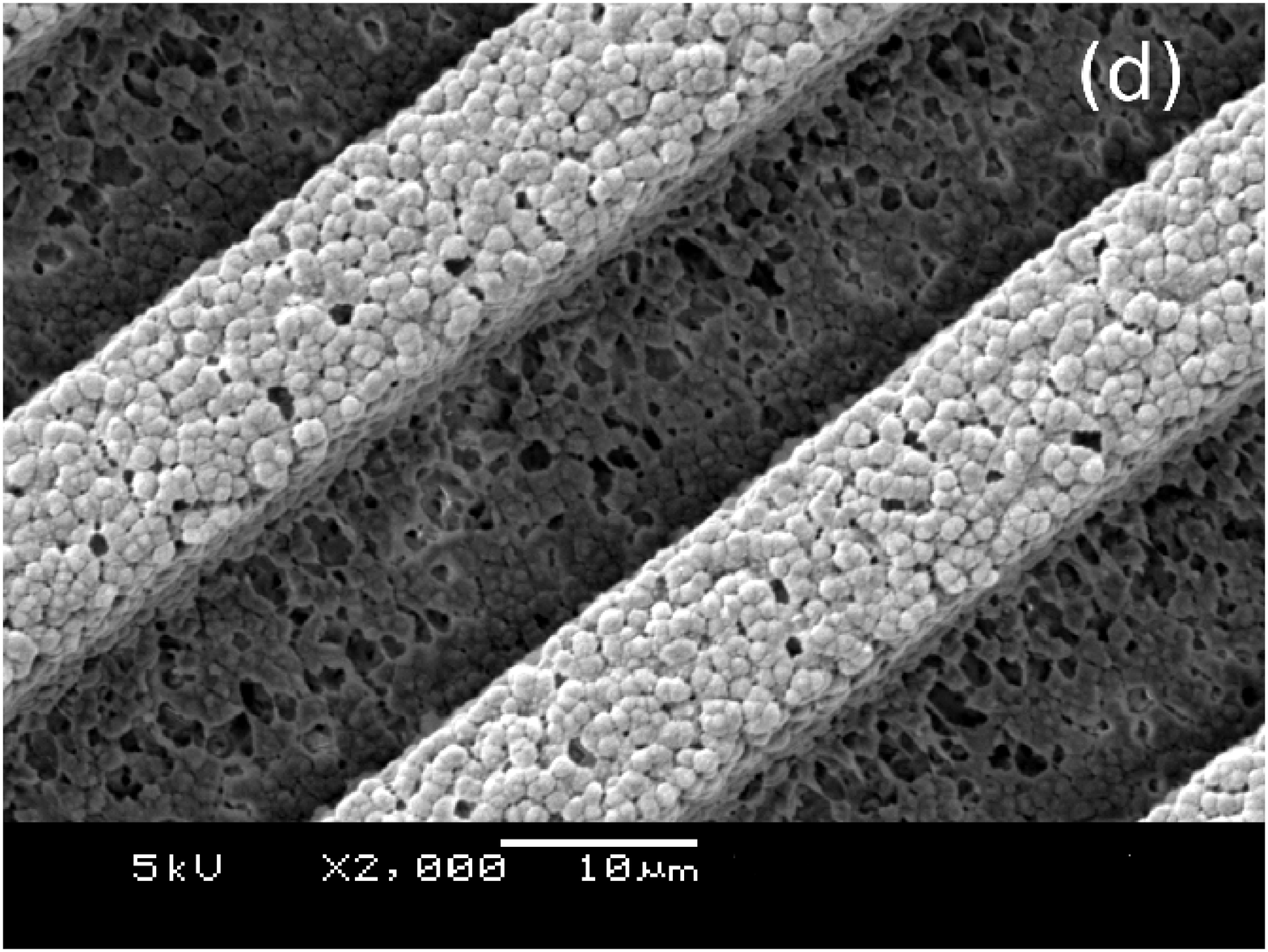}
\caption{\label{sem_pic_microchannels}Representative scanning electron microscope images of the used microchannels: (a) a smooth, flat PDMS microchannel of height $H = 50~\unit{\mu m}$ and of width $D = 320~\unit{\mu m}$; (b) a structured PDMS microchannel possessing longitudinal grooves of height $h = 20~\unit{\mu m}$ and of width $w = 16~\unit{\mu m}$; (c) a rough, porous PVDF rectangular microchannel of $H = 37~\unit{\mu m}$ in height, $D = 270~\unit{\mu m}$ in width, and $l = 30 ~\unit{mm}$ in length; (d) a patterned PVDF microchannel with grooves of $w = 15 \unit{\mu m}$, $L = 27~\unit{\mu m}$, and $h = 14~\unit{\mu m}$, showing its porosity and additional surface roughness.}
\end{center}
\end{figure}

Fig.~\ref{sem_pic_microchannels} shows representative scanning electron microscope images of the polymeric microchannels.  The used polymers are smooth PDMS (Polydimethylsiloxance, RTV 615 rubber component A and curing agent B, GE Bayer Silicones) and rough PVDF (Polyvinylidene fluoride, Hylar 460, Ausimont). 
These microchannels were fabricated with the micro-molding technique~\cite{micro-molding} by casting polymeric films on a wafer of desirable micropatterns.

The PDMS films were obtained by mixing the rubber component A with the curing agent B ($10:1$ by weight). The mixture was degassed and then poured onto the mold ($\sim 1~mm$) and cured in an oven for $3$ hours at $80^{o}$C.
For the second type of the channel, porous PVDF films were prepared via the liquid induced phase inversion technique~\cite{phase-separation1} to create microporosity at the surface.  We followed the recipe in Ref.~\cite{making_PVDF} to produce microporous PVDF membranes. A solution of $20~wt\%$ (by weight)  PVDF in NMP (1-methyl-2-pyrrolidinone, 99\% extra pure, Acros) was prepared by mixing with a mechanical stirrer for $12$ hours at $70^{o}$C. The viscous solution was cast on a mold with a thickness of $250~\mu m$ and immediately submerged in a $50:50~wt\%$ water/NMP bath for $30 - 60$ minutes. If the coagulation bath is lightly agitated the membrane releases itself from the mold after a few minutes. To remove most of the NMP from the membrane, it was submerged in water and ethanol (proanalysis grade, Merck) subsequently for $30 -60$ minutes each. The film was then taped to a piece of paper to prevent from curling-up and left to dry in the fume hood (1 hour) before placing it in a $30^{o}$C vacuum oven overnight.

The casting mold of the flat rectangular channel has the geometric dimension of length $l = 40~\unit{mm}$, height $H = 50~\unit{\mu m}$, width $D = 320~\unit{\mu m}$ (Fig.~\ref{setup}b schematically showing the resulting channel). Some molds containing longitudinal micro-grooves, as schematically shown in Fig.~\ref{setup}c, are employed with the additional microtextures of height $h = 20~\unit{\mu m}$ and of the varying width $w~=~8,~16,~32~\unit{\mu m}$, while keeping $w_s = w$. Here, the width of the microstructure unit is $L=w+w_s$, with $w_s$ the width of the solid-liquid interface (i.e., the bottom part of the micro-groove). 
The dimensions of the mold are precisely replicated by the PDMS. However, the final dimensions of the PVDF channels and the grooves are smaller than those on the mold due to shrinkage in the phase separation and drying processes. This polymer shrinkage can be $25-32~\%$ and the exact channel dimensions are determined by the SEM measurements.

Unlike smooth PDMS, PVDF possesses micron and submicron interconnecting voids (shown in Fig.~\ref{sem_pic_microchannels} d) as a result of the phase separation process. Its surface pore size distribution is quite narrow with an average pore size of $1.8~\mu m~\pm~0.4~\mu m$, determined from SEM pictures. Both polymers are hydrophobic with similar static contact angles, measured to be $120^{o}$ for flat PDMS and $127.4^{o} \pm 2.6^{o}$ for flat porous PVDF (OCA 20, Dataphysics).
These large contact angles can help microstructured PDMS and PVDF surfaces sustain composite-wetting Cassie-Baxter (CB) state with air pockets underneath a still water droplet~\cite{Sbragaglia_PRL2007,Pirat_epl_2008}.  
From our experiences, sometimes the non-wetting state can be maintained, i.e., the case of apparent slippage, for our imposed flow rates. However, sometimes we observed that the non-wetting state is unstable and turns to a wetting state in the course of the experiments.  What causes this transition is not so clear because the employed microstructures are chemically homogeneous so is the wettibility across the channels. Presumably the dissolution of the trapped gas may introduce the transition.  During experiments, we frequently checked if the CB state was maintained since a transition to Wenzel state can take place.

The enclosure of the microchannels was a cover glass of thickness $170~\mu m$ (D263 borosilicate, Louwers).  The glass slide was treated with oxygen plasma (Plasmafab 508, Electrotech) for $30$ seconds at $100 W$ and then bound to the hydrophobic polymer channels.  The flow inlet and outlet were connected through the polymer substrates. The microfluidic channel was placed on and controlled by a high precision positioning stage (New Focus) with an angle adjustment to ensure the horizontality of the microfluidic assembly with respect to the optical axis of the objective lens. 

\subsection{Details of the $\mu$-PIV technique}

Following the standard working principle of the first $\mu$-PIV method~\cite{1stPIV}, our setup is similar to that in Ref.~\cite{CPirat_Lab_Chip08}. The fluid consists of degassed milli-Q water (vacuum pumped at $\approx 10$ mbar for $\sim 3$ hours) seeded with $300~nm$ diameter fluorescent polystyrene particles (R300, Duke Scientific Corp.), with a peak excitation wavelength $542~nm$ (green light) and a peak emission wavelength $612$ nm (red light). The seeding particles have the mass density of $1.05$ g/cm$^3$ close to be neutrally buoyant tracers that faithfully follow the flow field. 
The channel illumination is produced by a dual cavity diode-pumped Nd:YLF laser at $527 nm$ (Pegasus-PIV, New Wave Research Co) via an oil immersion Plan-Apochromat $100\times$ objective with numerical aperture NA $=1.4$ (Carl Zeiss), mounted in an inverted microscope (Carl Zeiss, Axiovert 40 CFL).  The microspheres absorb green laser light and emit red light, which passes through a dicromatic mirror to a recording system.  With a proper delay time $\Delta t$ between two exposures, image pairs were recorded by a cooled sensitive double shutter CCD camera with the resolution $1376 \times 1040$ pixels $\times~12$ bit (Sensicam qe double shutter, PCO. Imaging).  Typical $\Delta t$ in our experiments ranged from $1~ms$ to $25~ms$ depending on the flow rate.  The precise focusing in the desired plane of the microchannel was controlled by an accurate piezoelectric objective-lens positioning system (MIPOS 500, Piezosystem Jena GmbH) with the finest step of $100~nm$.

We discuss various errors associated with \mPIV technique below.  The depth of field $\delta_z$ for a microscope objective lens limits the accuracy of the determination of the reference base plane ($z = 0$), marking the interface between liquid and the solid microridges~(as the reference dash line in Fig.~\ref{setup} c). It is estimated with the following equation~\cite{1stPIV,1stPIV_1999}:
\begin{equation}
\delta_z = \frac{n_o\lambda}{{\rm NA}^2}+\frac{n_o e}{M{\rm NA}} \approx 0.6~\mu m,
\end{equation}
where $n_o=1.518$ is the refractive index of the immersion medium of oil used for the objective lens, $\lambda$ is the wavelength of the particle emission light, NA is the numerical aperture of the objective lens~(our effective NA $\approx 1.3$ for the oil-immersion objective viewing the water channel), $e$ is the pixel resolution of the CCD camera or a single pixel of the CCD camera; 
and $M$ is the total magnification of the system.  The error in determining the reference plane $z = 0$ was approximately $0.5~\mu m$ although the reproducibility of piezoelectric driven positioning system is less than $100~nm$.  Because of the volume illumination used in $\mu$-PIV, out-of-focus particles unfortunately could contribute to correlation signals.

Clusters of particles, even when they are out-of-focus, can sufficiently contribute to the correlation signals and hence lower the accuracy of the results.  To eliminate this error, the space-averaged intensity of each interrogation window was calculated and compared to the space-time-averaged one, set as a reference.  Prior to cross-correlation, interrogation windows of intensity larger than the reference one by $2\%$ were removed; on average $\sim 20\%$ of the total interrogation windows were not used.  Reducing the influence of cluster particles gave some improvement of the analysis.  As a result, our data have high signal-to-noise ratios so we did not apply any velocity validation and smoothing to the results.

The interrogation view of $\mu$-PIV images was about $120~\mu m \times 127 \mu m$.  When processing data, small interrogation windows of the typical size of $64 \times 16$ pixels ($\sim 8 \times 2~\mu m$) were used to achieve high spatial resolution for resolving detailed velocity fields.  Interrogation windows were overlapped by $50\%$ to satisfy the Nyquiest sampling criterion~\cite{1stPIV_1999}.  Each velocity field in a plane at the different channel height $z$ was processed from $150-200$ image pairs so as to reduce the random vectors due to Brownian thermal motion.  In addition, velocity profiles were improved by an ensemble-averaging correlation method~\cite{1stPIV_1999}. This technique sequentially includes cross-correlating particle image fields for each image pairs, ensemble-averaging the resulting cross-correlation functions, and determining the peak of the ensemble-averaged correlation function. This peak represents the averaged displacement of particles in one interrogation window; consequently, with the given time interval ($\Delta t$) the flow velocity is determined. 
The flow in the experiment is homogeneous and steady downstream in $\hat{x}$ so in general the velocity profile we present below is a streamwise average over the probing interrogation window; we denote $\overline{u_x} \equiv \langle u_x \rangle_x$.

\section{Results and Discussion}

\subsection{The laminar velocity profile over flat hydrophobic surface}

\begin{figure}
\begin{center}
\includegraphics[width=3.6in]{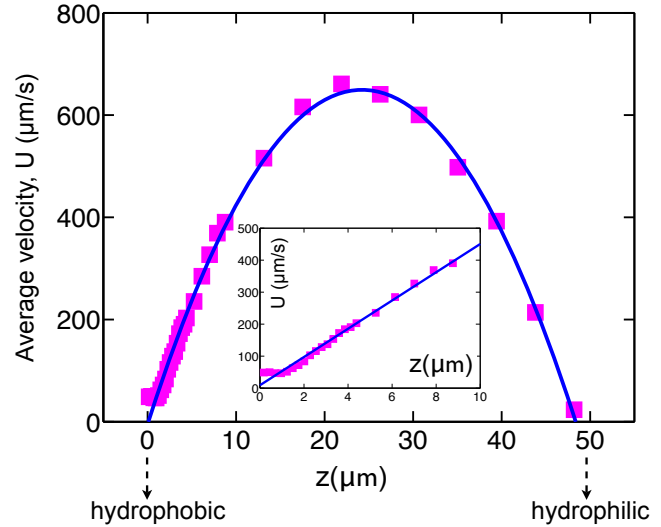}
\caption{\label{V_PDMS_flat}Average velocity profile over a smooth rectangular PDMS microchannel (Fig.~\ref{sem_pic_microchannels}a). The solid line is a quadratic fit, showing a parabolic profile in the laminar flow.  The inset shows the detailed velocity profile close to the hydrophobic PDMS surface; the solid line is the best linear fit. 
}
\end{center}
\end{figure}

Fig.~\ref{V_PDMS_flat} reveals an overall average velocity profile $U=\langle u_x\rangle_{x,y}$ streamwise as a function of $z$ past a flat hydrophobic PDMS substrate ($z=0$) which is enclosed by a thin hydrophilic glass at the plane close to $z = 50~\mu m$.  The overall velocity profile exhibits a parabolic profile--indicated by the solid line as the best quadratic fit for the laminar flow profile.  The fit shows the expected result that the highest flow velocity appears in the central plane of $z = 25 \mu m$.

The inset in Fig.~\ref{V_PDMS_flat} displays a linear relation between velocity $U$ and $z$ close to the hydrophobic flat wall for $z \apprle 10 \mu m$.  The corresponding slip length $b$, determined by the ratio of the velocity to the velocity gradient at the boundary, can be calculated based on Navier's condition:
\begin{equation}
U(z=0)= b~\partial U/ \partial z |_{z=0}\label{Eqn4b},
\end{equation}
where $U$ is the averaged tangential velocity and the boundary $z=0$ marks the solid-liquid interface. 
We estimated the slip length $b$ from the bulk velocity profile for $z \apprle 10 \mu m$, using the best linear fit shown in the inset, and found $b = 200 \pm 200~nm$.  The error bar here is calculated from different experimental \mPIV measurements with new PDMS samples.

Closely looking at the inset in Fig.~\ref{V_PDMS_flat}, a plateau of the velocity profile is noticeable in the vicinity of the hydrophobic wall $z  \apprle 1 \mu m$.  This type of plateau has been observed in similar type of measurements in microfluidics~\cite{Olga_PRE_2003, PTabeling_PRE_2005}, and may be attributed to the falsely represented flow velocity by the velocity of the seeding particles. Particle density in PIV has been observed to be non-uniformly distributed; that is the tracking particles tend to accumulate in the place of high flow velocity. 
In the vicinity of the wall, seeding particles may experience hydrodynamic interaction due to the wall or electrostatic forces within the electric double layer ~\cite{PTabeling_PRE_2005}. 
As discussed in Ref.~\cite{Olga_PRE_2003}, the microspheres can be affected by the Taylor dispersion~\cite{GITaylor}, i.e., large tangential shear right next to the wall enhances the effective diffusivity of the tracers and thus increases their migration speed.
Note that the particles outside this plateau regime were not affected by these influences and do represent the true flow velocity.
Despite the plateau area, our estimate of the slip length is the commonly used method of the extrapolation from the bulk velocity profile ($z \apprle~10~\mu m$ in our calculation) so as to lessen the weight of the velocity measurement right next to the wall.

The slip lengths over various hydrophobic flat substrates have been investigated with other, either direct or indirect, techniques, for instance, surface force apparatus, atomic force microscope, near-field laser velocimetry, double  and molecular dynamic simulations (see Ref.~\cite{ELauga_noslip_BC_2007} and the references therein).  The up-to-date convincing measurements reach an agreeable value of $\approx 20~nm$~\cite{Craig_review_Rep_Prog_Phys05, Bocquet_PRL2006_2}.  Unfortunately, the resolution of standard \mPIV is limited due to the volume illumination via a magnifying objective of large NA. For conventional $\mu$-PIV, the best achievable resolution is in the order of submicrons~\cite{PTabeling_PRE_2005}, one order of magnitude larger compared to the resolution achieved by atomic scale apparatus.  Very recent advances in \mPIV using evanescent waves and nanoscaled particles can reach $0.01~\mu m$ resolution for nanofluidic study~\cite{PTabeling_PRL_2008}.  Nevertheless, in general the conventional $\mu$-PIV offers a well-established flow probe for microscopic fluid dynamical systems on the micrometer and sub-micron scales.

\subsection{The detailed velocity profiles over microstructured hydrophobic surfaces}

\begin{figure}
\begin{center}
\includegraphics[width=3.6in]{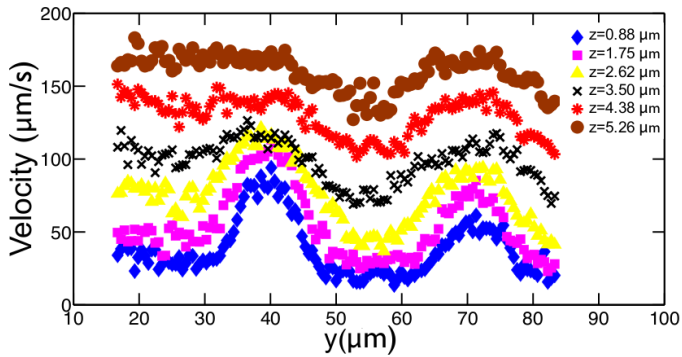}
\caption{\label{mPIV_microstructrues_PDMS}Detailed streamwise velocity profiles $\overline{u_x}(y)$ at different channel heights $z$ when water flows over a micropatterned PDMS surface (as shown in Fig.~\ref{sem_pic_microchannels}b).}
\end{center}
\end{figure}

Additional roughness provided by hydrophobic micro-grooves (Fig.~\ref{setup}c) can sustain the CB state, offering a fraction of the shear free boundary condition, and thus increase the flow velocity.  Fig.~\ref{mPIV_microstructrues_PDMS} shows such velocity profiles of $\overline{u_x}$ in a microstructured PDMS channel of $w = 16 \mu m,~h = 20 \mu m, D = 320 \mu m$, and $H = 50 \mu m$.  These velocity profiles reveal higher velocity regimes for $34 \mu m \apprle y \apprle 50 \mu m$ and $66 \mu m \apprle y \apprle 82 \mu m$ where the water is flowing over the air-liquid interfaces.  Conversely, lower velocities--showing large drag--appear above the solid-liquid interfaces where the no-slip boundary condition holds.   

\begin{figure}
\begin{center}
\includegraphics[width=3.5in]{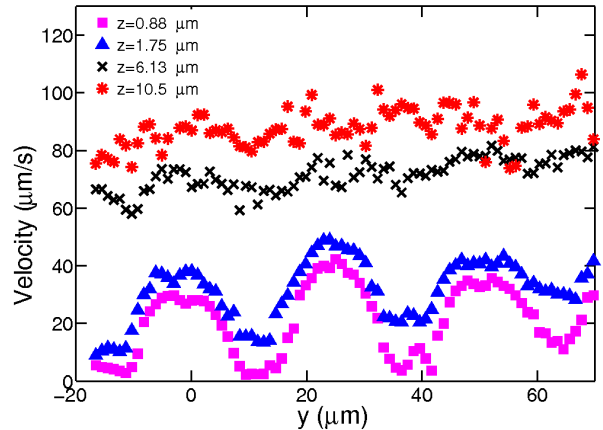}
\caption{\label{mPIV_microstructrues_PVDF}$\mu$-PIV measurements of the detailed velocity profile over a microstructured porous PVDF substrate, as shown in \ref{sem_pic_microchannels}d. }
\end{center}
\end{figure}

Fig.~\ref{mPIV_microstructrues_PVDF} reveals the \mPIV measurements of $\overline{u_x}$ at various channel heights $z$ over the porous PVDF surface as shown in Fig.~\ref{sem_pic_microchannels}c. The velocity profiles close to the hydrophobic wall show sinusoidal features reflecting the variation of local boundary conditions.  The shear-free air-liquid interfaces enhance the flow velocities in comparison to no-slip liquid-solid areas. Nevertheless, the influence of varying boundary conditions is restricted since the oscillatory feature of the velocity is not observed for $z > 10 \mu m$.

From the \mPIV measurements, such as Figs. \ref{mPIV_microstructrues_PDMS} and \ref{mPIV_microstructrues_PVDF}, the slip length $b$ can be readily calculated from the average velocity profile $U(z)$. By assuming Navier's condition in Eqn.~\ref{Eqn4b}, we extracted the effective slip length from the bulk velocity profile with a best linear fit close to the wall for $z  \apprle 10 \mu m$. 
Two methods were used to characterize the slip length. One represents the local effective slip length, $b_{l}$, determined by the averaged velocity $U$ merely over the air-liquid interface; the other is the global effective slip length, $b_{g}$, calculated with the averaged velocity $U$ over a unit cell which includes one solid-liquid and one air-liquid area.

\subsection{The geometric effect of micro-grooves on slip length}

\begin{figure}
\begin{center}
\includegraphics[width=3.3in]{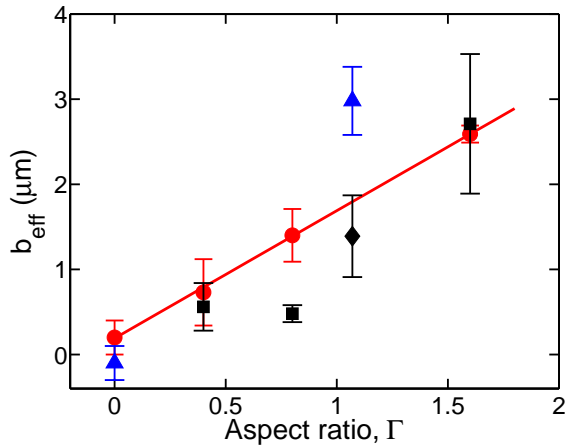}
\caption{\label{slip_length_all}Effective slip length {\it vs.} aspect ratio, $\Gamma=w/h$, over various hydrophobic microstructured surfaces.  Data are obtained by velocity profiles averaged over different areas: (1) the local effective slip length $b_l$ over the air-liquid interfaces of PDMS ($\bullet$) and PVDF ($\blacktriangle$), with the line showing the best linear fit of $b_l$ for PDMS ($\bullet$), and (2) the global effective slip length $b_g$ over the combination of the air-liquid and liquid-solid interfaces of PDMS ($\blacksquare$) and PVDF ($\blacklozenge$).} 
\end{center}
\end{figure}

Fig.~\ref{slip_length_all} shows the slip lengths, $b_l$ and $b_g$, as function of the aspect ratio $\Gamma$ of micro-grooves; $\Gamma=w/h$. The data for $\Gamma =0$ represents the effective slip length for the microchannels without patterns (Fig.~\ref{setup}b), so in this case $b_l = b_g$. 
The local effective slip length $b_l$, extracted from the averaged velocity profile over the air-liquid interface, clearly shows a linear increase as $\Gamma$ increases (shown by the symbol $\bullet$ with the best linear fit).  The error bars cover the scatter of slip lengths calculated from various experiments.  
These findings imply a proportional increase of slippage with increasing shear-free area.  Due to the friction at the solid-liquid interfaces, the global effective slip lengths $b_g$ show lower values comparing to $b_l$. Consistent with $b_l$, $b_g$ shows a similar trend when $\Gamma$ is increased, as shown in Fig.~\ref{slip_length_all}.  It is interesting to note that the slip length over hydrophobic microstructured porous PVDF surface has a relatively large value: $b_l \sim 3~\mu m$.  This suggests that the porous structure of PVDF can enhance slippage by either rough surfaces or extra voids that can sustain air.  To elucidate and disentangle these effects, simplified materials, for instance, some flat but porous hydrophobic surfaces with well-defined porosity, can be employed in $\mu$-PIV.

\begin{figure}
\begin{center}
\includegraphics[width=3.5in]{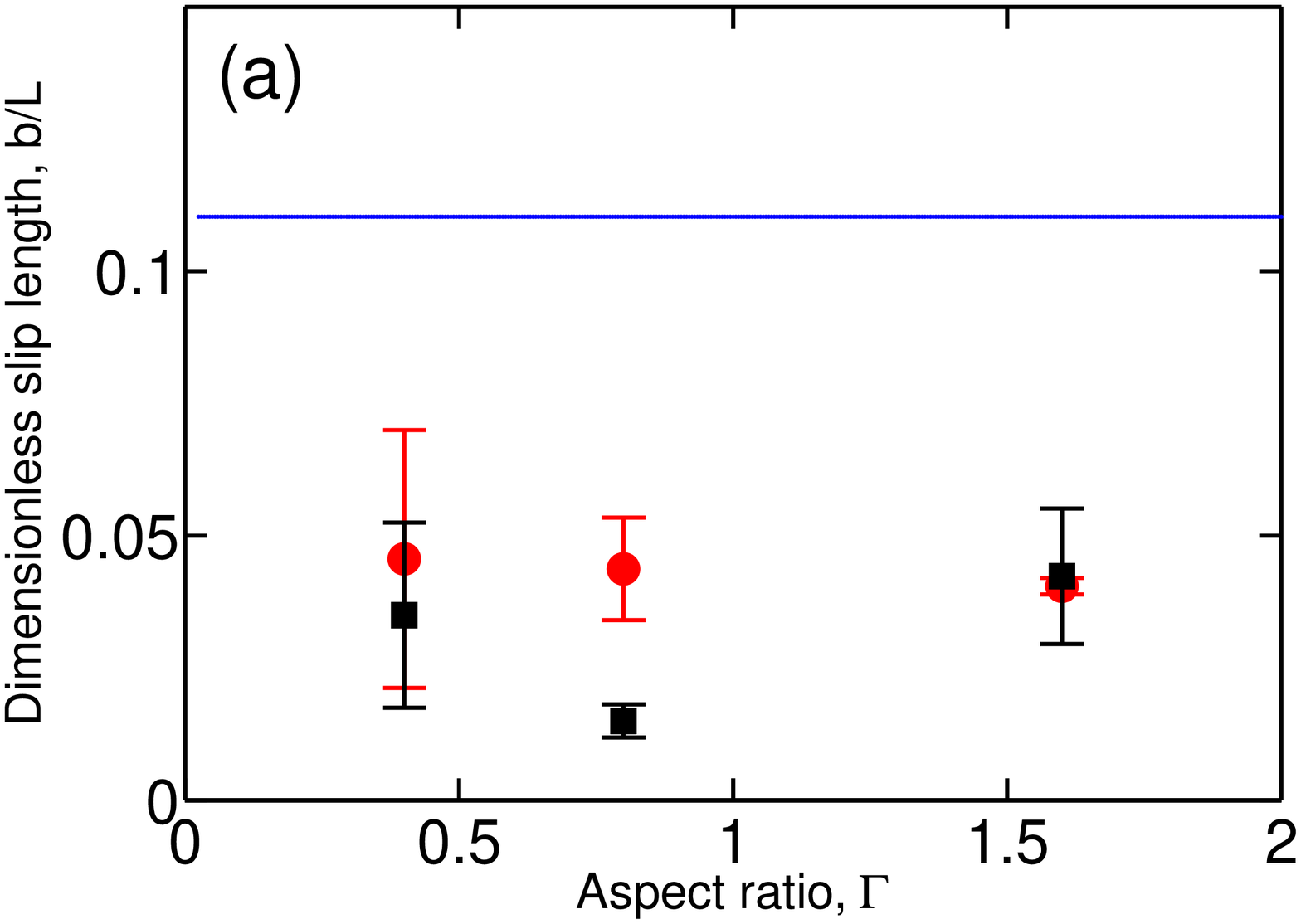}
\includegraphics[width=2.4in]{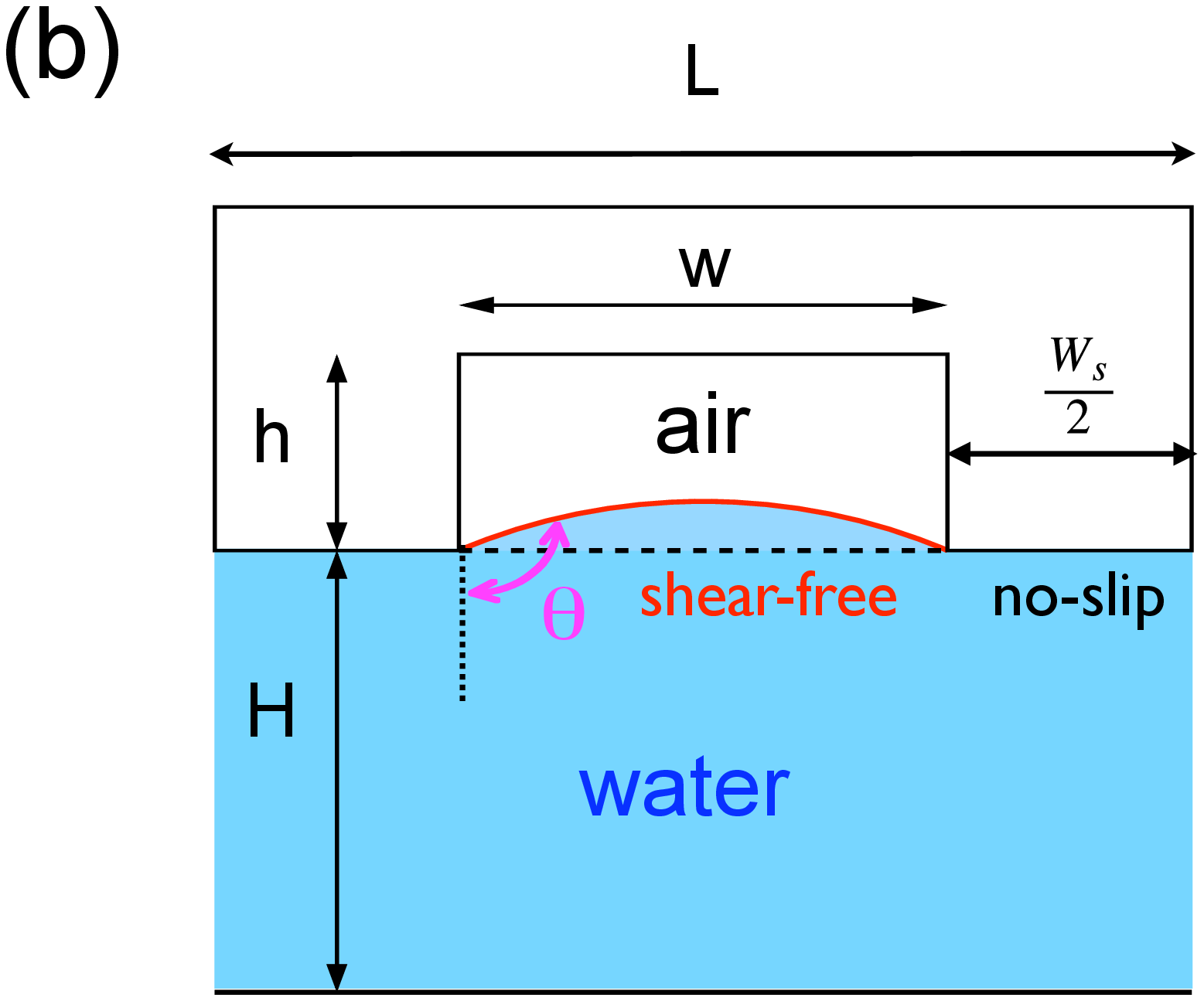}
\caption{\label{b_exp_vs_theory} (a) Dimensionless slip length, $b/L$, {\it vs.} aspect ratio $\Gamma=w/h$ over various microstructured channels of PDMS (the same data, $b_l$ ($\bullet$) and $b_g$ ($\blacksquare$), as shown in Fig.~\ref{slip_length_all}). The solid line is the theoretical prediction which is described in Eqn.~\ref{model_b}~\cite{JPhilip}. (b) A sketch of a bending air-liquid meniscus, complying with the wetting angle $\theta$ in a microstructured channel section.}
\includegraphics[width=3.2in]{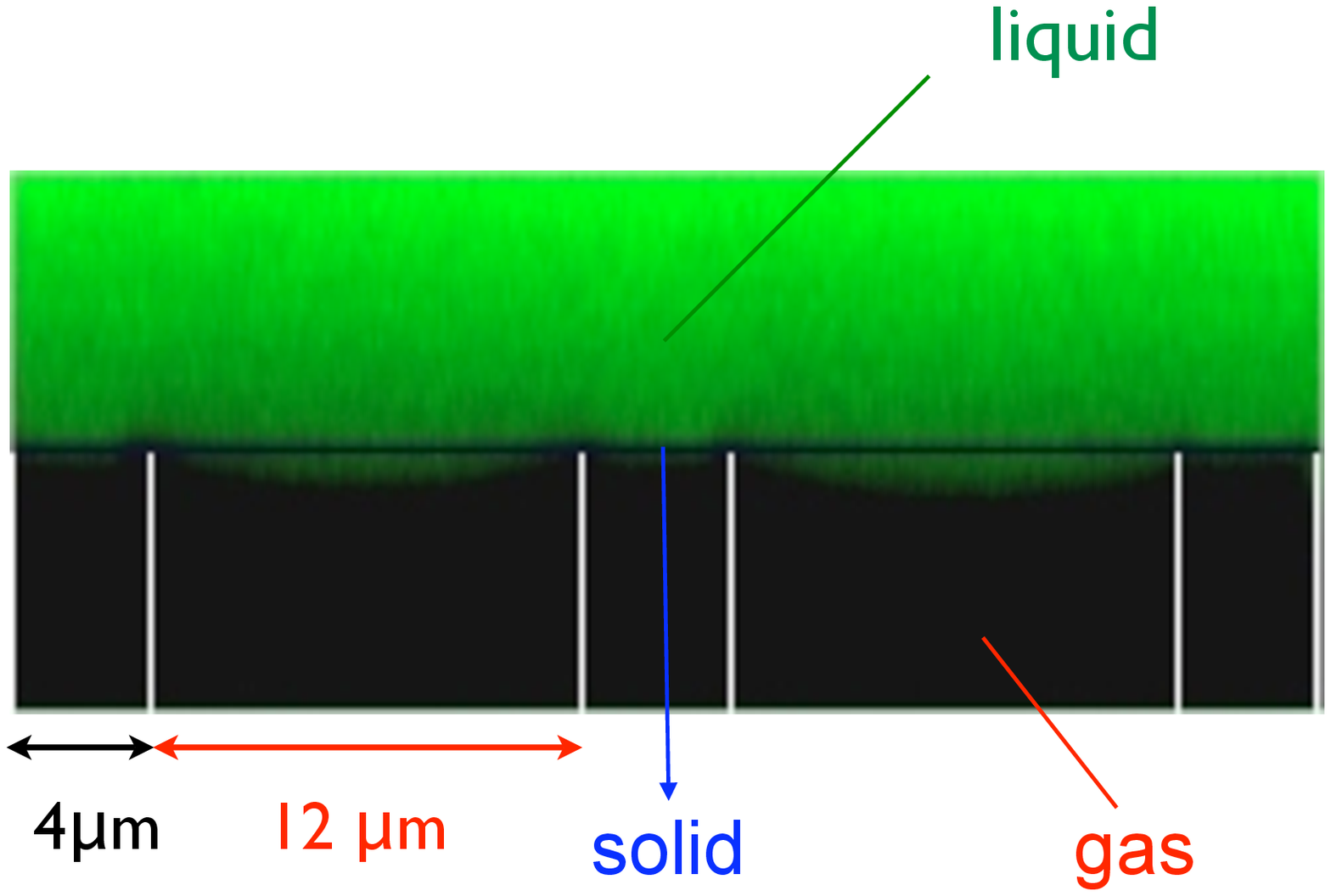}
\caption{\label{confocal_image}A confocal microscope image showing bending liquid-gas menisci for water passing through a micropatterned channel with a combination of the liquid-solid interfaces of the width of $4~\mu m$ and the liquid-gas interfaces of the width of $12~\mu m$.}
\end{center}
\end{figure}

To better understand our results, in Fig.~\ref{b_exp_vs_theory}a, we compare our measurements of slip lengths shown in Fig.~\ref{slip_length_all} with the  analytical result by Philp et al.~\cite{JPhilip} based on a mixed boundary condition of periodically alternating no-slip and shear-free boundary conditions on a smooth surface, namely 
\begin{equation}\label{model_b}
\frac{b}{L}=\frac{1}{\pi}\ln(1/\cos(\frac{\pi}{2}\frac{L-w}{L})),
\end{equation}
in a dimensionless form in which $b$ is normalized by the periodicity length $L$.
The analytical calculation, depicted by the solid line in Fig.~\ref{b_exp_vs_theory}a, shows a constant value of $b/L$ for a given fixed value of the shear free fraction, $w/L$.
However, this theoretical model assumes (i) an infinite tall channel, i.e., $H \gg h$, and $H \gg L$ and (ii) a flat shear-free interface.  Our measurements of $b$ shown in Fig.~\ref{b_exp_vs_theory} are systematically lower than the prediction.  The discrepancy can be explained by the confinement effect and the influence of a bending meniscus.  As discussed in Ref.~\cite{Mauro}, the slip length slightly decreases as the confinement of the entire no-slip wall of height $H$ is comparable to $L$.   The $H/L$ in our data ranges from $0.78$ to $3.13$.  In this range, a small decrease $\sim 0.01$ in $b/L$ is predicted in Ref.~\cite{Mauro}.  Moreover, a large decrease in $b/L$ has been predicted for the case of curved air-liquid interfaces~\cite{Mauro, LBocquet_EuroPhysJE04}.  This correction is proportional to the depth of the bending meniscus~\cite{Mauro}.  In our experimental parameter range, this additional decrease in $b/L$ as predicted in Fig. 4 in Ref.~\cite{Mauro} can span between $0.01$ and $0.06$, which may explain the lower values obtained in the experiments.  It is worth noticing that the protrusion of meniscus could be as large as a quarter of the groove-width $w$ conforming to the wetting contact angle $\theta = 120^{o}$ for flat PDMS (Fig.~\ref{b_exp_vs_theory}b).  Following this Young's angle to assume the meniscus depth of $w/4$, the calculation with Eqn.~45 in Ref.~\cite{Mauro} predicts a decrease of $0.03$ in $b/L$ for our experimental parameter of the shear-free fraction of $w/L=1/2$ for PDMS.  This decrease in $b/L$ can explain the gap in our comparison and thus suggests the crucial role of the bending menisci.

Fig. ~\ref{confocal_image} also confirms bending menisci between liquid-gas interfaces in an independent experiment viewed with a confocal microscope.  The seeded fluorescein in the water shows green light which presents not only in the channel but also protrude in the liquid-gas regime of the width of $12~\mu m$.

\begin{figure}[t]
\begin{center}
\includegraphics[width=3.4in]{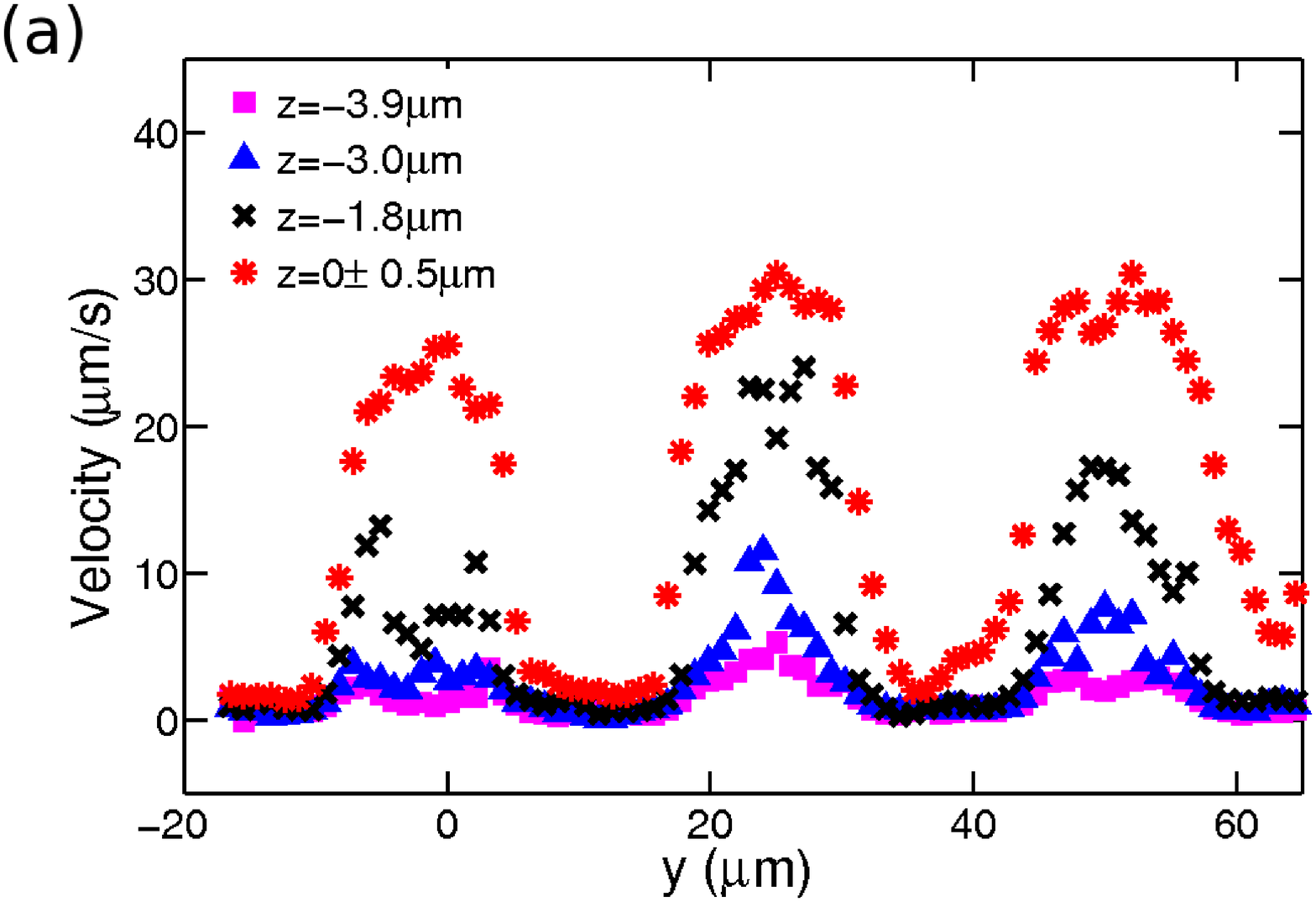}
\includegraphics[width=3.4in]{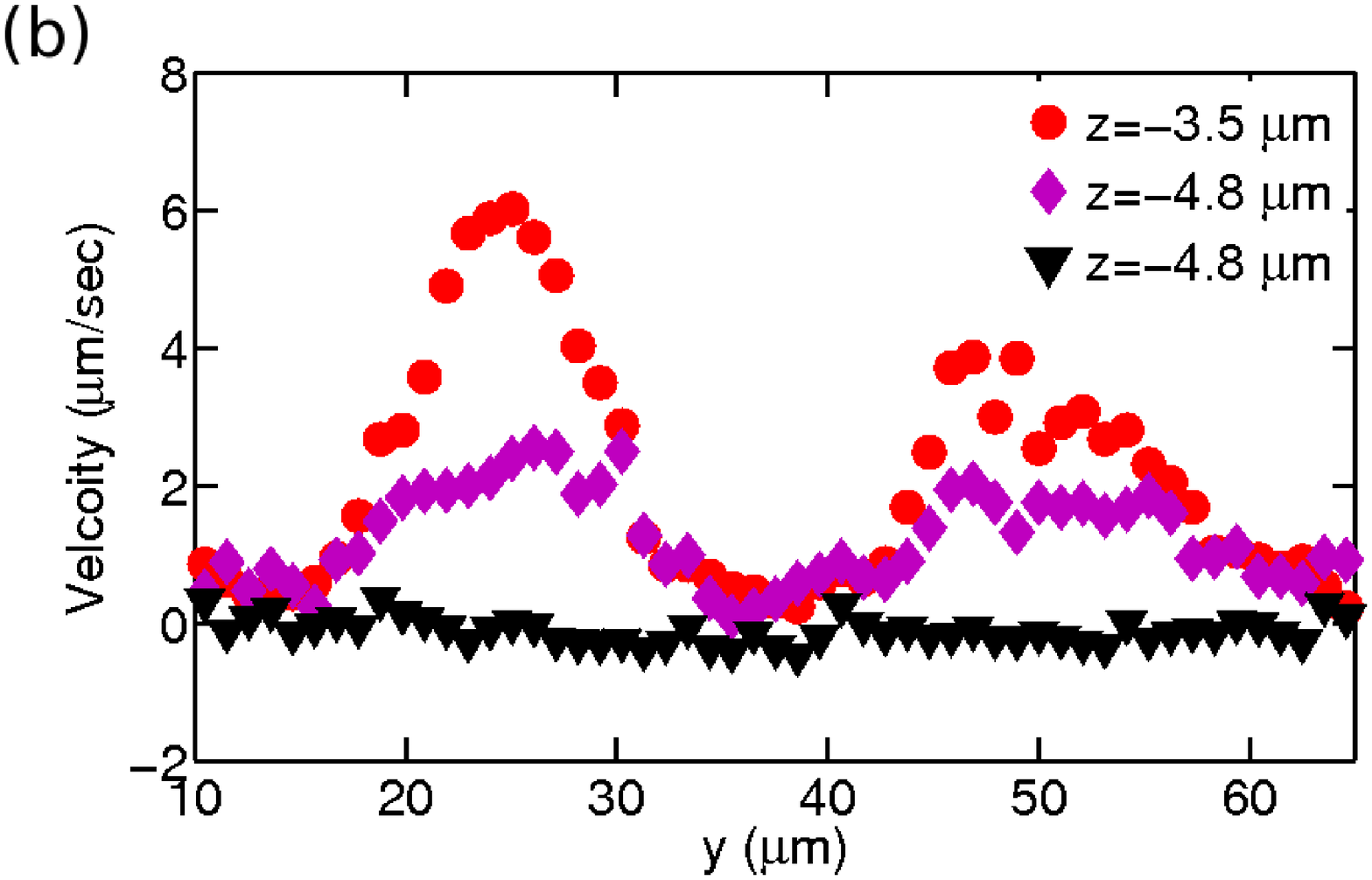}
\caption{\label{bending_meniscus}\mPIV measurements of the velocity profile inside PVDF micro-grooves below the base line $z=0$, showing curved menisci towards trapped air. In (a), different symbols indicate the velocity profile $\overline{u_x}(y)$ at different scanned planes: [$z=0$] ($\ast$) at the solid-liquid edge of the microridges, $z~=~-1.8~\mu m$ ($\times$), $z~=~-3.0~\mu m$ ($\blacktriangle$), and $z~=~-3.9~\mu m$ ($\blacksquare$).  (b) The zoomed-in velocity profile obtained from focused microspheres even deeper inside the same micro-grooves as in (a), revealing the longitudinal velocity $\overline{u_x}$ in the planes of $z~=~-3.5~\mu m$~($\bullet$) and $z~=~-4.8~\mu m$~($\blacklozenge$), and the transverse velocity $\overline{u_y}$ in the same plane of $z~=~-4.8~\mu m$~ ($\blacktriangledown$).} 
\end{center} 
\end{figure}

To study the influence of the bending of the meniscus, we also performed \mPIV inside the microridges, i.e., $z < 0$.  We observed curved air-liquid menisci protruding from the flat surfaces marked by the solid-liquid interfaces.   
Fig.~\ref{bending_meniscus}a shows the velocity profiles of the seeding spheres inside of the micro-grooves below the base plane $z=0$. The high velocity areas reflect the shear-free air-liquid menisci, while the essentially zero velocities indicate the rigid solid places.  In Fig.~\ref{bending_meniscus}b, the focused seeding particles still contribute to $\mu$-PIV measurements in the plane of $z = -4.8~\mu m$; a small velocity variation, within $3~\unit{\mu m/s}$ streamwise, is still detectable due to different local boundary conditions, whereas the transverse velocity $\overline{u_y}$ is essentially zero at $z = -4.8~\mu m$. These measurements imply the important role of the shape of the meniscus.

\section{Conclusions}
Different effective slip lengths of a few micrometers have been determined when water steadily flows over a variety of hydrophobic micropatterned surfaces.  Our experiments confirm the presence of the trapped air between hydrophobic microridges.  The air-liquid interface provides a drag-reducing mechanism with its shear-free boundary condition.  The slip length $b$ is enhanced as the width of the micro-grooves $w$ is increased due to the broader shear-free areas. This increase is consistent with a theoretical prediction, which uses a model of a mixed boundary condition consisting of periodically alternating shear-free and no-slip conditions.  However, our measurements show smaller slip length in comparison with the analytical calculation that assumes an infinite tall cell and flat air-liquid interfaces. 
This discrepancy can be explained by the confinement effect and the bending of air-liquid menisci.  Our \mPIV measurements performed inside micro-grooves show the protrusion of the menisci towards the trapped air, in contrast to the theoretical assumption of flat air-liquid menisci.  
This work presents experimental observation of a linear relation between the effective slip length and the aspect ratio of the micro-grooves over micropatterned hydrophobic surfaces with $\mu$-PIV. We also show \mPIV measurements of flow velocity due to the bending menisci between the micro-grooves. Our findings reveal the crucial role of the exact shape of the menisci in drag reduction for water flowing past micropatterned hydrophobic surfaces. So far, the experiments have been carried out with different aspect ratio $\Gamma$ by varying $w$.  Further intriguing investigations include experiments performed with different shear fractions and a broad range of $\Gamma$, with a variation of $h$, to shed light on the broader influences of meniscus geometry upon the slip length in micro- and nano- flows.

\begin{acknowledgments}
We gratefully thank M. Sbragaglia for stimulating discussion and Ineke P{\"u}nt for the preparation of the polymeric films and the help with the SEM pictures.
\end{acknowledgments}

\bibliography{slip_lengths_via_mPIV}

\end{document}